\documentclass[12pt,preprint]{aastex}
\shorttitle{UV Spectroscopy of Old Novae} 
\shortauthors{} 

\begin{document}

\title{Far Ultraviolet Spectroscopy of Old Novae II: RR Pic, V533 Her and DI Lac  
\altaffilmark{1}}

\author{Edward M. Sion, Patrick Godon, \& Liam Jones} 

\affil{Department of Astrophysics \& Planetary Science, Villanova University, \\ 
800 Lancaster Avenue, Villanova, PA 19085, USA}
\email{edward.sion@villanova.edu ; patrick.godon@villanova.edu ; 
ljones15@villanova.edu}



\altaffiltext{1}{Based on observations made with the NASA-CNES-CSA 
{\it Far Ultraviolet Spectroscopic Explorer (FUSE)}. {\it FUSE} was operated 
for NASA by the Johns Hopkins University under NASA contract 
NAS5-32985. }

\begin{abstract}
The old novae V533 Her (Nova Her 1963), DI Lac (Nova Lac 1910) and 
RR Pic (Nova Pic 1891) are in (or near) their quiescent stage 
following their nova explosions and continue to accrete at a high 
rate in the aftermath of their explosions. 
They exhibit 
continua that are steeply rising into the FUV as well as absorption 
lines and emission lines of uncertain origin. All three have FUSE 
spectra which offer not only higher spectral resolution but also 
wavelength coverage extending down to the Lyman Limit. For DI Lac, 
we have matched these FUSE spectra with existing archival 
IUE spectral coverage to broaden the FUV wavelength coverage. 
We adopted the newly determined interstellar reddening corrections 
of Selvelli and Gilmozzi (2013). 
The dereddened FUV spectra have been modeled with our grids of 
optically thick accretion disks and hot, NLTE white dwarf photospheres. 

The results of our modeling analysis  indicate that the hot 
component in RR Pic and V533 Her is likely to be the accretion disk
with a mass accretion of $10^{-8}M_{\odot}$/yr and $10^{-9}M_{\odot}$/yr
respectively. However, the disk cannot produce the observed absorption lines.
For the WD to be the source of the absorption lines in these two systems, 
it  must be very hot with a radius several times its expected size
(since the WD in these systems is massive, it has a smaller radius).  
For DI Lac we find the best fit to be a disk with 
$\dot{M}=10^{-10}M_{\odot}$/yr with a 30,000K WD. 

\end{abstract}

\section{Introduction}

Cataclysmic variables (CVs) are compact binaries comprised of a Roche-lobe filling sun-like star transferring gas to a white dwarf (WD) via an accretion disk (if the WD is non-magnetic) or via a magnetically channeled accretion column if the WD is magnetic. When a critical mass of hydrogen-rich gas accumulates on the white dwarf, an explosive thermonuclear runaway is triggered identified as a classical nova (CN).

Following the nova explosion, the stellar remnant enters a phase of 
non-explosive evolution during quiescence. This phase is not well 
understood. Do these post-novae evolve into dwarf novae as their 
accretion rates drop? When does nuclear burning and hence the soft 
X-ray production stop following the nova explosion? The time scale 
for this to occur is predicted to be in the range of 1 year to $>$ 10 
years, possibly as long as 300 years depending upon the mass of the 
WD and the amount of H left on the WD after the nova explosion, and 
is inversely dependent on the white dwarf mass. 
What are the accretion 
rates of old novae as a function of time since the nova explosion? 
Is there enhanced mass transfer due to irradiation of the secondary 
donor star (causing the donor to bloat and/or by driving a wind 
off of the donor star) by the hot white dwarf and/or hot accretion 
disk? Does the mass of the white dwarf grow following a nova outburst? 

An essential key to answering a number of these questions is in the far ultraviolet 
wavelength domain where the Planckian peak in the energy distribution 
of a Cataclysmic variable accretion disk or accreting white dwarf occurs. Specifically, a reliable determination of the rate of accretion onto each post-nova white dwarf and a comparison of the accretion rates of each system, at different times since their last eruption could bear directly on the accretion related questions mentioned above.

The old novae RR Pic, V533 Her and DI Lac have good quality far 
ultraviolet spectra on the MAST archive obtained with the Far 
Ultraviolet Spectroscopic Explorer (FUSE), and International Ultraviolet 
Explorer (IUE). 
In addition, archival FUSE spectra enables our model fitting to 
extend down to the Lyman Limit which provides additional insights 
and  constraints on the nature of the hot emitting component. 
Moreover, RR Pic, offers a number of advantages for a synthetic spectral 
analysis. RR Pic itself has an accurate trigonometric parallax and 
very good  FUSE spectra. Our analysis 
utilizes the new Hubble FGS parallax distance of 400 pc which removes 
one critical free parameter in our model fitting. The physical and 
orbital parameters from the literature for RR Pic, V533 Her and 
DI Lac are summarized in Table 1.

\clearpage

\begin{deluxetable}{ccccccccc}  
\tablewidth{0pc} 
\tablecaption{
        Physical and Orbital Properties of Old Novae 
}
\tablehead{
System    & Nova    & $P_{orb}$ &  i        &           d     &   V      & E(B-V)   & $M_{\rm wd}$ &  Speed Class  \\ 
Name      &  Year   & hr        & deg       &  pc             &          &          & $M_{\odot}$  &           
} 
\startdata 
RR Pic   & 1925$^a$ & 3.481$^b$ & 65$^{cf}$ & 380-490$^c$     & 12.3$^d$ & 0.00$^a$ &    0.95$^e$  &  Slow$^c$     \\ 
V533 Her & 1963$^a$ & 3.53$^f$      &  62$^g$       & 560-1250$^i$ $^j$    &          & 0.03$^a$ &    0.95$^g$      &   Slow$^d$  \\ 
DI Lac   & 1910$^a$ & 13.6$^m$  & $<$18$^i$     &           &          & 0.26$^a$ &              &                                  
\enddata 
\tablenotetext{  
a}{Selvelli, P. \& Gilmozzi, R. 2013, A\&A, 560, 49
}  
\tablenotetext{  
b}{Vogt, N. 1975, A\&A, 41, 15.
}  
\tablenotetext{  
c}{Schmidtobreick, L., Tappert, C. \& Saviane, I. 2008, MNRAS, 389, 1348
}  
\tablenotetext{  
d}{Warner, B. 1985, Mon. Not. R. ast. Soc., 219.
}  
\tablenotetext{  
e}{Haefner, R., Metz, K. 1982, A\&A 109, 171.
}
\tablenotetext{  
f}{McQuillin et al. 2012, MNRAS, 419, 330
}
\tablenotetext{  
g}{Rodriguez-Gil and Martinez-Pais 2002        }  
\tablenotetext{  
h}{Kraft, R. 1964, ApJ, 139, 457
}  
\tablenotetext{  
i}{Slavin et al.1995, MNRAS, 276, 353 
} 
\tablenotetext{  
j}{Gill, C. D. \& O'Brien, T. J., 2000, MNRAS, 314, 175
} 
\end{deluxetable} 

\clearpage 

In section 2, we present a log of the FUSE and IUE spectroscopic 
observations and describe the spectral line features in the FUSE 
wavelength range. In section 3, we present the results of a search 
for line and continuum variations as a function of orbital phase. 
In section 4, we describe our synthetic spectral fitting codes, 
model grids and fitting procedure, while in section 5, we present 
our model fitting results. Finally, in section 6, we summarize our 
conclusions.   

\section{\bf Far Ultraviolet Spectroscopic Observations}

The observing log of far ultraviolet spectroscopic observations  
of RR Pic, V533 Her and DI Lac is presented in Table 2. All spectra were downloaded from the MAST archive

All of the FUSE spectra of the three systems were acquired through the LWRS aperture in TIME-TAG mode.  Each exposure was made up of multiple subexposures. As pointed out by Godon et al. (2012), the FUSE reduction 
requires a careful post-processing of the different FUSE spectral segments
before they are co-added into one final spectrum. 
Extensive details on the acquisition and the processing of the FUSE data 
are given in Godon et al. (2012) and will not be repeated here.

The FUSE spectra for the three systems along with their respective line identifications are displayed for RR Pic in Fig. 1, V533 Her in Fig.2 and DI Lac in Fig.3. The absorption line and emission line identifications for the FUSE (as well as for IUE) spectra of RR Pic, V533 Her and DI Lac are comprehensively tabulated by Selvelli \& Gilmozzi (2013). 

In the FUSE wavelength range, RR Pic has a rich absorption line spectrum 
as noted by Selvelli and Gilmozzi (2013) who tabulate equivalent 
widths (EW) and Full Widths at Half-Maximum (FWHM).  
Limiting our focus to the strongest absorption features 
(EW $>$ 0.75\AA), they are Lyman Epsilon (938), S\,{\sc vi} (934 \& 945), 
Lyman Delta (950), S\,{\sc iii} (1012 \& 1016), S\,{\sc iii} (1063, 1073), 
P\,{\sc v} (1118), blended P\,{\sc v} (1128.38) +
Si\,{\sc iv}  (1128.3) and C\,{\sc iii}  (1175).

For V533 Her, all of the emission lines in its FUSE spectrum are sharp and 
of geocoronal origin with full widths at half-maximum (FWHM) of a 
fraction of an Angstrom. However, V533 Her, like RR Pic, exhibits a 
rich absorption line spectrum. The strongest features (EW $>$ 0.75\AA) 
are due to S\,{\sc iv} (1063), Si\,{\sc ii} (1108), P\,{\sc v} (1118), 
Si\,{\sc iv} (1123), P\,{\sc v} (1128), Si\,{\sc iv} (1128), and 
C\,{\sc iii} (1175). 
The strongest among these features are the P V lines. 
If the abundances derived from these P lines are suprasolar, then 
their overabundance points to explosive CNO burning and hence are 
asssociated with the 1963 and earlier nova explosions of V533 Her 
(cf. Sion and Sparks (2015)).

\begin{deluxetable}{cccccc} 
\tablecaption{Observation Log}  
\tablehead{ 
System & Telescope &  Data ID    & Obs. Date  & Time     & Exp.Time    \\ 
Name   &           &             & YYYY-MM-DD & hh:mm:ss &   sec                   
}
\startdata 
RR Pic &  FUSE     & D9131601001 & 2003-10-29 & 13:35:41 & 1792      \\                
       &  FUSE     & D9131601002 & 2003-10-29 & 14:40:16 & 3916   \\ 
       &  FUSE     & D9131601003 & 2003-10-29 & 16:24:04 & 3733   \\ 
       &  FUSE     & D9131601004 & 2003-10-29 & 18:07:39 & 4185   \\ 
       &  IUE      & SWP05775    & 1979-07-11 & 23:15:54 & 1200  \\  
DI Lac &  FUSE     & D9131301001 & 2003-10-20 & 09:38:42 & 2572       \\ 
       &  FUSE     & D9131301002 & 2003-10-20 & 11:02:47 & 3473  \\ 
       &  IUE      & SWP29325    & 1986-09-28 & 18:09:18 & 16799 \\  
V533 Her & FUSE    & D9131701001 & 2003-07-01 & 03:34:36 & 548    \\ 
         & FUSE    & D9131701002 & 2003-07-01 & 04:19:26 & 1390    \\ 
         & FUSE    & D9131701003 & 2003-07-01 & 05:15:22 & 496   \\ 
         & FUSE    & D9131701004 & 2003-07-01 & 05:59:29 & 1687  \\   
       &  IUE      & SWP44805    & 1992-05-29 & 07:52:12 & 18000 \\  
\enddata
\end{deluxetable} 

\clearpage

\clearpage 

For DI Lac, unlike RR Pic and V533 Her,the FUSE spectrum does not clearly 
reveal strong absorption features. The spectrum is rather noisy and, like 
V533 Her, it is strongly affected by ISM molecular hydrogen absorption.  

The IUE spectra of the 3 old novae are briefly displayed together on a log-log
scale in Fig.4. 
The clear signature of wind outflow is seen in the IUE spectra of
DI Lac and V533 Her spectra with the C IV (1550) resonance doublet 
present with P Cygni structure. The presence of this feature with its blue-shifted absorption along with the blue-shifts of other absorption features points to a wind from the disk and/or boundary layer.
We will come back to description of the IUE spectra of the 3 systems 
in the results section.  

\section{Phase Dependent FUV Line and Continuum Variations}

In order to shed light on the origin of the absorption lines in the FUSE spectra of the three systems, we examined the subexposures that were obtained over each individual FUSE orbit. For RR Pic, there were four such exposures, for DI Lac two exposures and for V533 Her, four individual exposures. 
The details (timing) of each spectrum are listed in Table 3.

For RR Pic, each subexposure is essentially one orbit of the FUSE spacecraft. 
The average spacecraft orbital period is 90 minutes. The first exposure started 
on Oct. 29 at 13:35:41, and the last exposure started on Oct. 29 at 18:07:39, 
and lasted for approximately 4000s. Hence, RR Pic was observed with FUSE for 
approximately 6 hours, compared with its orbital period of 3.481 hours.

The spectra obtained in orbit 1 and 3 are almost identical to each other and 
have a higher flux level than the spectra obtained in orbit 2 and 4, which are 
themselves almost identical to each other.
We have co-added exposures 1 and 3 together, as well as 2 and 4. 
In Fig.5, we have overplotted the two resulting co-added spectra.  

The absorption line features are slightly blue-shifted for exposures 1 and 3 and are 
slightly red-shifted for exposures 2 and 4. This is consistent with the occulting 
material coming from the L1 stream over-shooting the disk, because during that phase, 
the WD is moving away from the observor, therefore explaining the red-shift observed 
in exposures 2 and 4 with less flux (in a manner similar to the FUSE observations 
of EM Cyg, Godon et al. 2009). However, the material absorbing the flux does 
not change the depth of the lines but reduces the overall flux (luminosity) 
of the spectrum. Thus, the absorption lines do not form from the material 
overshooting the disk. In view of this line behavior, the possiblity remains 
that the absorption is associated with the accreting white dwarf itself.  
We note also that the IUE spectra of RR Pic exhibit the same drop of
flux at the orbital phase when the emission lines are slightly red-shifted 
compared to when the lines are slightly blu-shifted. The IUE SWP spectrum
of RR Pic that 
we retrieved from the archive was obtained when the flux was maximum,
i.e. at the orbital phase where the emission lines are blue-shifted.  

The total exposure time for V533 Her is 1.145 hrs, compared with its orbital period of 3.53 hrs. Our search for variations in the line and continuum flux proved more difficult than expected due to the more noisy subexposures.

For the fainter DI Lac, we have only two FUSE subexposures to examine. 
The total FUSE exposure time was 6045 s or 1.679 hrs, compared with its orbital period of 13.6 hrs. Hence, this amounts to a little over 12\% of its orbit. From our examination of the two subexposures, there is little to suggest any variation in the continuum level and absorption lines during the FUSE observation.

\section{\bf Synthetic Spectral Analysis}

We adopted model accretion disks from the solar composition 
optically thick, steady state disk 
model (``the standard disk model'') 
grid of Wade and Hubeny (1998). In these accretion disk models, the 
outermost disk radius, $R_{\rm out}$, is chosen so that 
$T_{\rm eff}$ ($R_{\rm out}$) is near 10,000K since disk annuli beyond this 
point, which are cooler zones with larger radii, would provide only a very 
small contribution to the mid and far UV disk flux, particularly in the 
FUSE and SWP FUV band pass. For the disk models, unless otherwise specified,
we selected every combination of $\dot{M}$, inclination and white dwarf 
mass to fit the data:  
the inclination angle i = 18, 41, 60, 75 and 81 
degrees, $M_{\rm wd}$ = 0.80, 1.03, 1.21 $M_{\odot}$ and 
$\log(\dot{M})$ ($M_{\odot}$/yr) = -8.0, -8.5, -9.0, -9.5, -10.0, -10.5. 
For the WD models, we used TLUSTY Version 203 (Hubney 1988) and 
Synspec48 (Hubeny and Lanz 1995) to construct a grid solar composition
WD stellar photospheres, with temperatures 
from 12,000K to 60,000K in steps of 1,000K to 5,000K, and with 
effective surface stellar gravity, Log(g),  
corresponding to the white dwarf mass of the accretion disk model. 
We adopt a projected standard stellar rotation rate $V_{\rm rot} 
\sin(i) $ of 200 km/s.  
We carried out synthetic spectral fitting with disks 
and photospheres, and a combination of both to model the FUSE spectra, 
the IUE spectra, and, when possible, 
the combination of the FUSE + IUE spectra to attempt to consistently 
fit a broader wavelength baseline.

\section{\bf Synthetic Spectral Fitting Results}

\subsection {RR Pic}

The IUE spectrum of RR Pic is consistent with  $\sim$15,000 K component 
and our synthetic spectral fits yield either a  
low WD surface temperature and/or low mass accretion rate,
giving an extremely short distance ($< 20$pc). 
The IUE spectrum of RR Pic is displayed in
Fig.4 together with standard disk models for comparison.  
{\it The slope of the continuum of 
the IUE spectrum of RR Pic is much more shallow than that of 
a standard disk model,} and could indicate the presence of 
a non-standard disk. 

On the other hand, however,
the FUSE spectrum of RR Pic reveals a rising FUV continuum 
(toward shorter wavelengths) and strong absorption features suggesting 
either a hot photosphere and/or a hot accretion disk. 
Therefore, we only model the FUSE spectrum of RR Pic.   
For the spectral fits (going down the Lyman limit) 
we have co-added the spectra from orbit 1 with that of orbit 3, 
when the source is least likely veiled by material flowing over the 
disk's rim.  The FUV slope of the FUSE continuum points to the presence 
of a hot component.  If we assume that the 
FUSE flux is due to a hot white dwarf, then we obtain a very high 
temperature of the order of $T_{\rm eff}$ = 70,000K to 80,000K 
with an inflated radius
of the order of $\sim 0.03 R_{\odot}$ ($\approx$20,000km) to account 
for the distance. 
Such a model is shown in Fig.6.
For $M_{\rm wd} = 1 M_{\odot}$ with a non-inflated radius, 
the scale factor-derived 
distance is reduced to $\sim$100 pc which is well below the parallax distance. 
 
For accretion disk model fits to the 
FUSE spectrum, with i = 60 degrees, $M_{\rm wd} = 0.8 M_{\odot}$ and a high accretion 
rate of $10^{-8} M_{\odot}$/yr, 
we obtain a distance of 427pc, which is within the 
range of RR Pic's parallax distance. 
This model has a flux level that is too low in the shorter wavelengths of FUSE.  
If we increase the WD mass to $1.03 M_{\odot}$, then the fit in the
shorter wavelengths is slightly improved and a distance of 506 pc is obtained.  
This accretion disk fit is shown in Fig.7. 
In order to fit the absorption lines with the disk model, 
the inclination of the disk has to be set to 3deg,
which is almost a face-on disk, while the inclination of RR Pic is 
known to be 65deg. 

Since the absorption lines do not arise 
in the disk, (see Fig.7), then they may arise from the white 
dwarf photosphere. If true, this would imply that we could be seeing
significant flux from the post-nova hot white dwarf photosphere itself. 
However, unless the WD is very hot with an inflated radius, 
the WD contributes too little flux for this to be a viable possibility: 
the addition of a WD (with a non-inflated radius) 
to the disk disk model does not improve the disk model.

\subsection{V533 Her}

The IUE spectra of V533 Her also reveal a relatively shallow slope of the 
continuum consistent with a cold component, namely, either a low mass 
accretion rate disk, or a low temperature WD, or a combination of both.
The IUE spectrum of V533 Her is also displayed in Fig.4 
in comparison to standard accretion disk models.  
The resulting distance we obtain from the low mass accretion
disk model fits, however, is far too close
and we are forced to reject these solutions.  
The slope of the IUE spectrum, on the overall,
agrees with that of a $\sim$15,000 K component.            

While this could be due to a non-standard disk, the FUSE spectrum 
of V533 Her has a slope  consistent with a hot FUV component with a
temperature  40,000K $\pm$5,000 K, the exact value depends on the
assumed WD surface gravity $\log(g)$. 
For a $1 M_{\odot}$ model, a hot white dwarf fit  
yields a distance that is too short (300-400pc), about 2-3 times smaller than 
the accepted value. Since the distance obtained from the model
scales like the WD radius, a possible solution would be a hot
WD with an radius inflated to 2-3 times its value.
In Fig.8 we present a solar abundances 40,000K WD fit with $\log(g)=8.6$, 
corresponding to a $1 M_{\odot}$ WD with a radius of 6,000km,
giving a distance of 316pc.  

Turning to accretion disk models, a disk around a 
$1 M_{\odot}$ WD with an accretion rate of $10^{-9} M_{\odot}$/yr 
and i = 60 degrees produces a good continuum fit and a reasonable 
distance of 832 pc.
This disk model is shown in Fig.9. Here too, the disk does not fit any
of the absorption lines.  

The disk+WD fit to the FUSE spectrum of V533 Her produces the 
best fit when each component contributes about half of the flux. 
This is achieved for $\dot{M}=10^{-9.5}M_{\odot}$/yr and 
$T_{\rm wd}=60,000$ K, giving a distance of 640pc and where the
WD contributes 51\% of the flux and the disk contributes the
remaining 49\%. 
This model is shown in Fig.10.  

Note, however, that while the model disk fits match the FUSE slope 
better, disk models clearly do not fit any of the absorption features. 
This is important since a hot white dwarf model does fit the absorption 
lines in the FUSE spectrum quite well. Nevertheless, we can virtually 
rule out the white dwarf because it contributes too little FUV flux,
unless it is very hot with an inflated radius two to four times larger
than the expected 6,000km. 

\subsection{DI Lac}

The distance we obtain from the model fit scales with the radius of the WD 
which itself depends on the WD mass. Since both the WD mass 
and distance to the system are unknown, and the spectra are of a low 
quality, there is a degenracy in the modeling. 
Therefore, we assume a standard WD mass  
of $M_{\rm wd} = 0.8M_{\odot}$, $\log(g)=8.4$, and an 
inclination of 18deg in the disk models.    

The IUE spectrum of DI Lac presents a moderately shallow continuum slope
consistent with a component with a temperature of about 20,000K.
A WD with T=20,000K gives an unrealistic distance of barely 40pc
(that would scale to less than 100pc assuming a low WD mass),  
while the disk model agreeing with the slope has a mass accretion rate of
$10^{-10}M_{\odot}$/yr and gives a distance of 140pc
(see Fig.4).  

The FUSE spectrum of DI Lac is consistent with a component
with a temperature of about 30-35,000K. A $0.8M_{\odot}$ WD with 
a temperature of 35,000K gives a distance of $\sim 150$pc, while 
a 30,000K WD gives a distance of $\sim 110$pc. 
An accretion disk model best fit to the FUSE spectrum 
has a mass accretion rate of  
$10^{-10}M_{\odot}$/yr (giving a distance of 140pc) 
to $3 \times 10^{-10}M_{\odot}$/yr (and a distance of 300pc). 

Since both the FUSE and IUE spectra of DI Lac are of pretty low quality 
but are consistent with each other as they 
give similar results, we decide to model the combined FUSE+IUE
spectrum. The best fit model to the combined FUSE+IUE spectrum of 
DI Lac is a disk + WD model, where a 30,000K WD helps fit the absorption 
lines in the FUSE range of the spectrum, while a disk with 
$\dot{M} = 10^{-10}M_{\odot}$/yr provides the lower flux needed to
fit the IUE range of the spectrum.  
The distance obtained from this model is d = 175pc.  
This combination  FUSE + IUE fit is displayed in Fig.11 showing the
FUSE range, and in Fig.12 showing the IUE range. 
The dotted line is the contribution of the white dwarf and the dashed 
line is the flux of the model accretion disk. 

\section{Summary}

It is apparent from our study of the hot components in the old novae 
addressed in this paper that the range of wavelengths extending down 
to the Lyman Limit by the FUSE spacecraft is essential to uncovering 
the nature of the hot coponent, be it a white dwarf, or an accretion 
disk. The FUSE coverage reveals that for RR Pic, the hot component 
is more likely represented by a bright accretion disk with the corresponding 
accretion rate of $10^{-8}M_{\odot}$/yr. An accretion rate this high 
is supported by optical observations of the accretion rates of old novae 
(e.g. Warner 1995 and references therein). This best-fitting accretion 
disk solution has a white dwarf of $\sim 1 M_{\odot}$, a disk 
inclination of 60 degrees and yielded a distance to RR Pic of 506 pc 
which agrees well with the parallax distance. 
We find that a hot WD ($T\sim 70-80,000$K) also fits the FUSE spectrum
of RR Pic, but implies a WD radius inflated to about 20,000km to 
agree with the observed distance. 

Likewise for V533 Her, 
the FUSE spectra provide fitting solutions with an accretion disk 
around a $1 M_{\odot}$ WD and a disk inclination of 60 degrees 
yielding an accretion rate of $10^{-9}M_{\odot}$/yr at distance of 
832 pc. But as in the case of RR Pic, the strong absorption features 
are not reproduced by the accretion disk.
Here too we cannot rule out a hot WD with an inflated radius 
as the source of the FUV: a 40,000K WD fits the FUSE spectrum 
and some of the absorption lines, assuming a WD radius of $\sim 12,000$km 
to 24,000km.  
A combined WD+disk model also yields a reasonable fit with a 
60,000K WD providing 51\% of the flux and a $10^{-9.5}M_{\odot}$/yr
disk providing the remaining 49\%, giving a distance of 640pc.  

For DI lac, we opt to combined the FUSE spectrum with the 
matching IUE spectrum which together provide self consistent results. 
We find that the best fit is obtained with a combined disk+WD model,
where the WD has a temperature of 30,000K.  
Our derived accretion rate of $10^{-10}M_{\odot}$ is 
lower by about an order of magnitude than both RR Pic and V533 Her.
Although Nova Lac 1910 is considerably earlier than Nova Her 1963, 
Nova Pic 1925 is almost as old as Nova Lac 1910 but has an accretion 
rate two orders of magnitude higher. 

The existing IUE spectra of RR Pic and V553 Her both exhibit a continuum
slope much more shallow than that of the standard disk model 
(see Fig.4) 
and in disagreement with the FUSE spectra revealing a hot component
that can be matched with an accretion disk.  
It is a possible indication that the standard disk model does not apply here,
and it also prevented the use of the IUE spectra in our spectral analysis. 

This work is supported by NASA ADP grants NNX13AF12G and NNX13AF11G 
to Villanova University. One of us, LJ, was supported by a Villanova 
Undergraduate Research Fellowship (VURF). This paper is based on 
observations made with the NASA-CNES-CSA Far Ultraviolet Spectroscopic 
Explorer (FUSE). FUSE was operated for NASA by the Johns Hopkins U
niversity under NASA contract NAS5-32985.

\begin{figure}
\vspace{-5.cm} 
\plotone{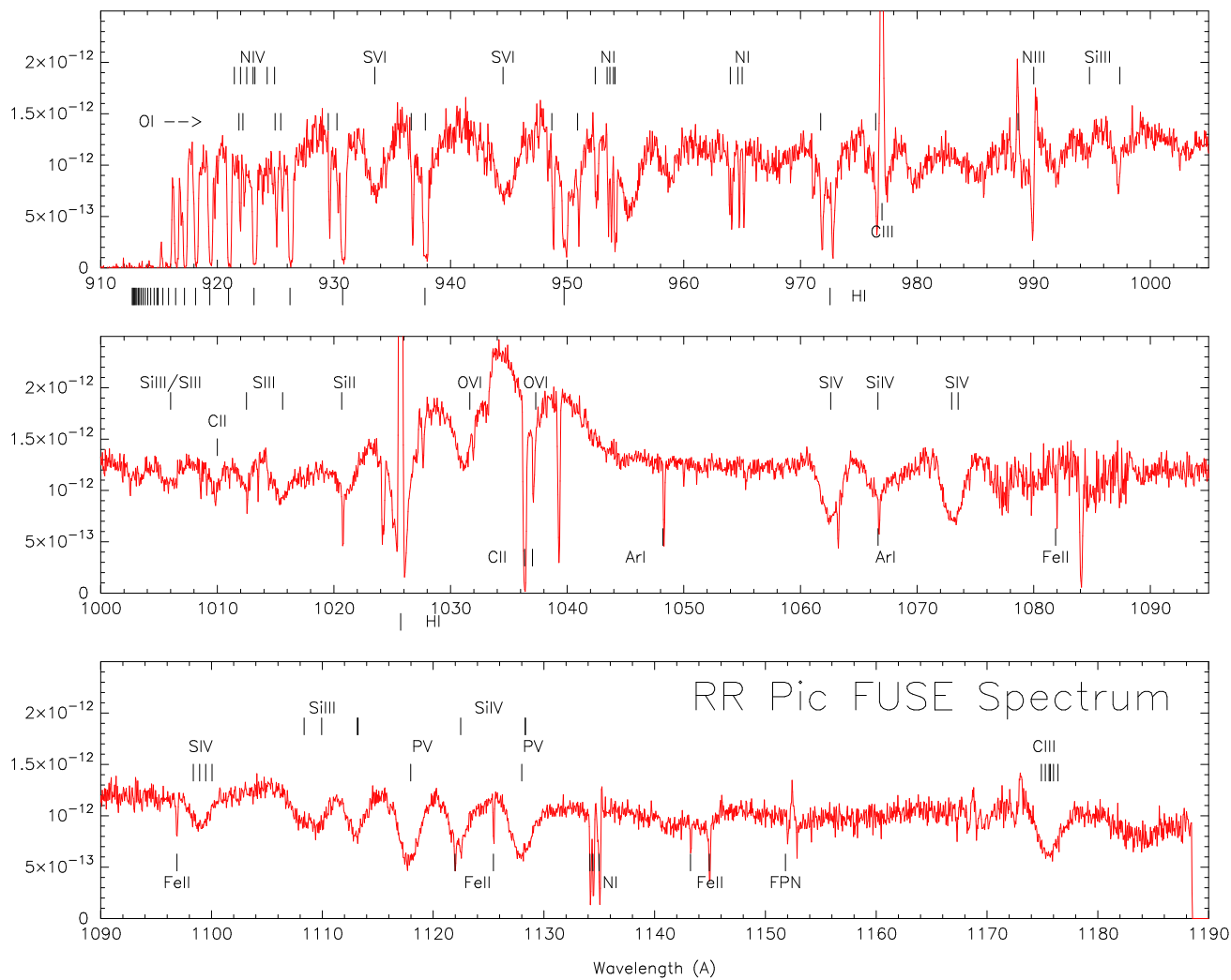} 
\vspace{1.cm} 
\caption{
The FUSE spectrum of the old nova RR Pic. 
The strongest identified line features are indicated by the tick marks. 
Most of the sharp absorption lines are probably  
from the interstellar medium, or from circumbinary material. 
Unlike, V553 Her and DI Lac, this spectrum does not show any 
ISM molecular hydrogen absorption line. 
The broad absorption lines are from the source, they consist mainly
in highly ionized species. The broad absorption around 955\AA\ 
is possibly the N\,{\sc iv} 955.33 line, though it is rarely observed 
in CVs and that region is often heavily affected by ISM molecular 
hydrogen absorption.  
The O\,{\sc vi} doublet exhibits a broad emission
feature and its broad absorption lines are blue shifted unlike the other 
broad absorption lines from the other species.  The sharp emission lines are due
to sunlight reflected in the telescope. A detector fixed pattern noise (FPN) is marked around 1052\AA .  
} 
\end{figure}

\begin{figure}
\vspace{-5.cm} 
\plotone{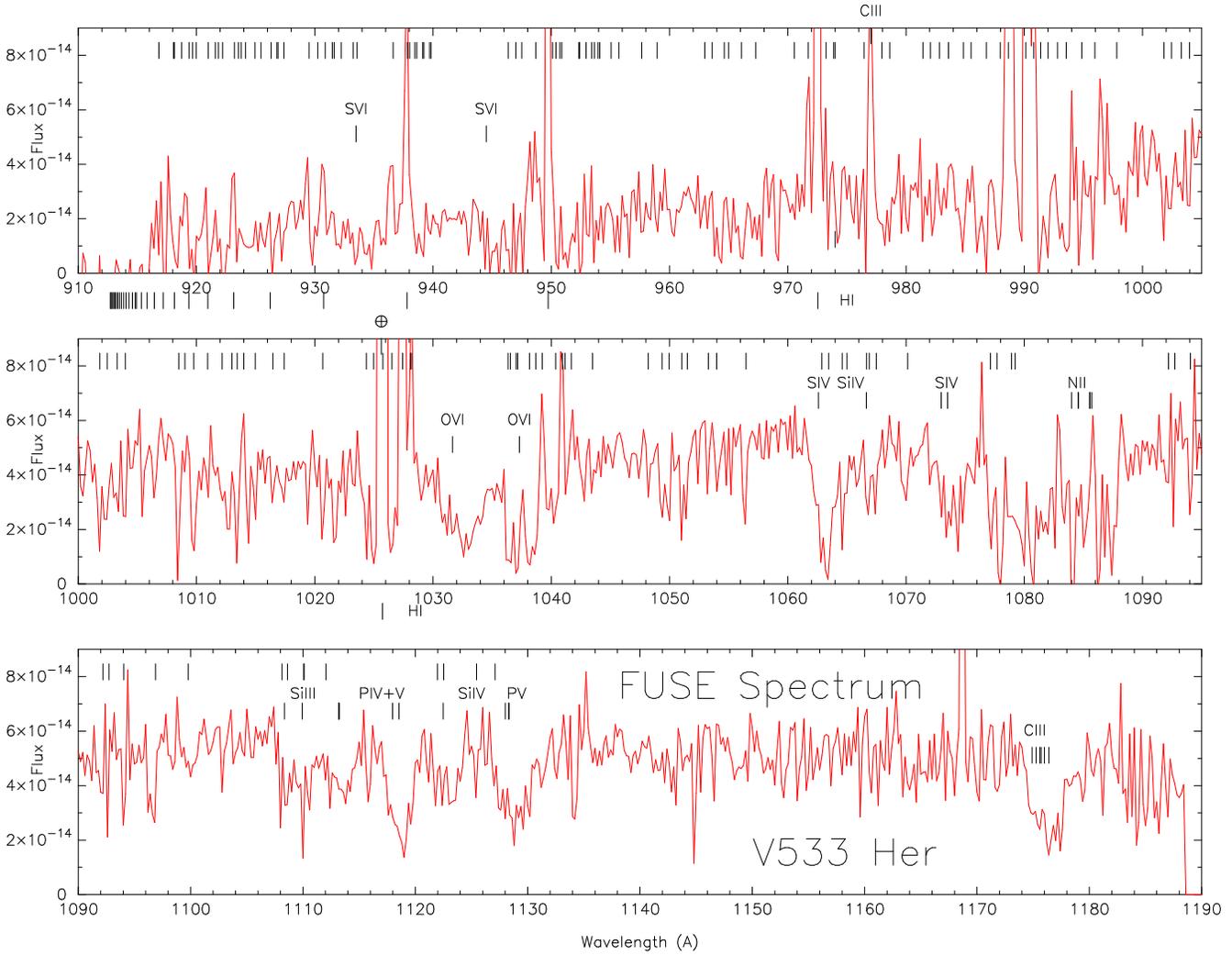} 
\caption{
The FUSE spectrum of the old nova V533 Her with line identification. 
This spectrum presents a multitude of ISM molecular hydrogen lines
that have been marked, together with ISM O\,{\sc i} lines, 
with vertical tick marks at the top of each panel.  
Here too, the sharp emission lines 
are helio-coronal in origin. Some broad high level ionization species
absorption lines have been identified with the source as marked on the figure.  
} 
\end{figure}

\begin{figure}
\vspace{-5.cm} 
\plotone{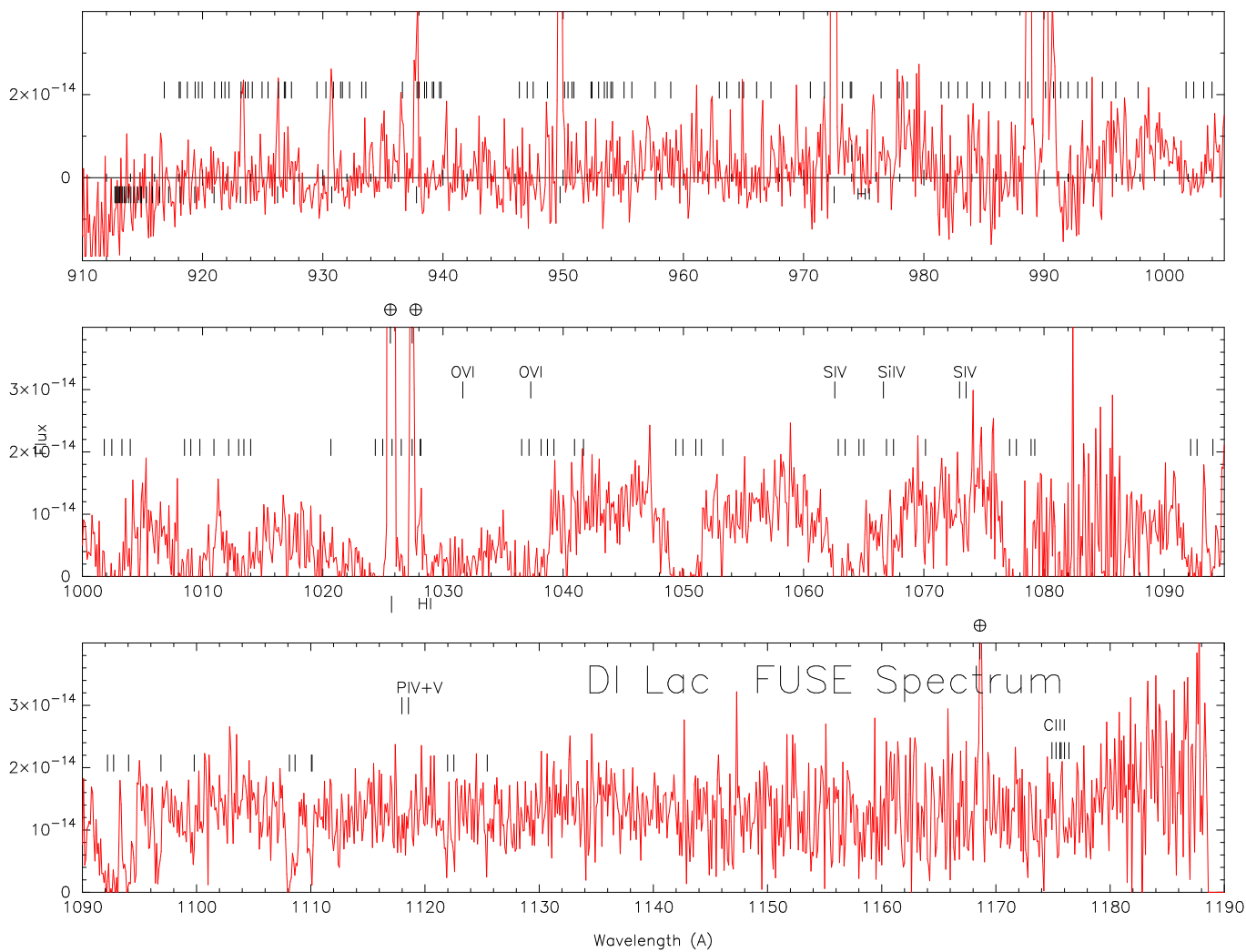} 
\caption{
The FUSE spectrum of the old nova DI Lac with line identifications. 
This spectrum is more noisy and it is strongly affected by ISM 
absorption lines (marked as in Fig.2), which literally slices the spectrum every 15\AA\ 
or so. 
Most of the spectrum in the shorter wavelengths ($< 1,000$ \AA ) 
is not reliable as it is dominated by sharp airglow emission lines 
and noise. We shifted  the upper panel upward to show where
the `positive' flux is of the same order as the `negative' flux. 
Some absorption lines from the source have been marked where they could be  
expected but they are not clearly identified. 
} 
\end{figure}

\begin{figure}
\vspace{-10.cm} 
\plotone{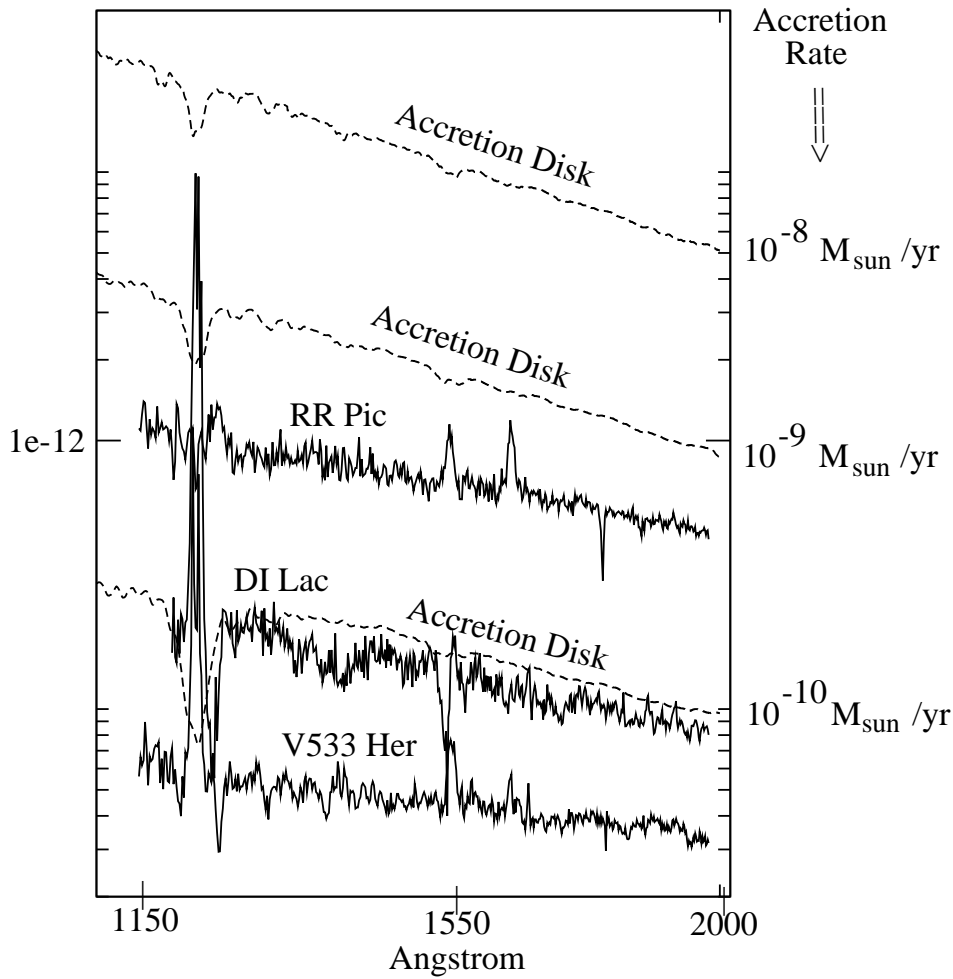} 
\caption{
The dereddened IUE SWP spectra of RR Pic, DI Lac and V533 Her are 
displayed, as indicated, on log-log scale. 
For comparison 3 standard accretion disk models have also been
displayed (dashed lines) for different mass accretion rates
as shown on the right hand side of the figure. 
The accreting white dwarf has mass of $1 M_{\odot}$ and 
the disk has an inclination of 60 degrees. The continuum slope of the 3 
IUE spectra is shallow and agrees with a low mass accretion
rate standard disk model ($\sim 10^{-10}M_{\odot}$/yr) or a 
relatively cold component. 
The high accretion rate standard disk models
($dot{M}=10^{-8}-10^{-9}M_{\odot}$/yr)
have a continuum slope steeper, indicative of a higher temperature.  
} 
\end{figure}

\begin{figure}
\vspace{-10.cm} 
\plotone{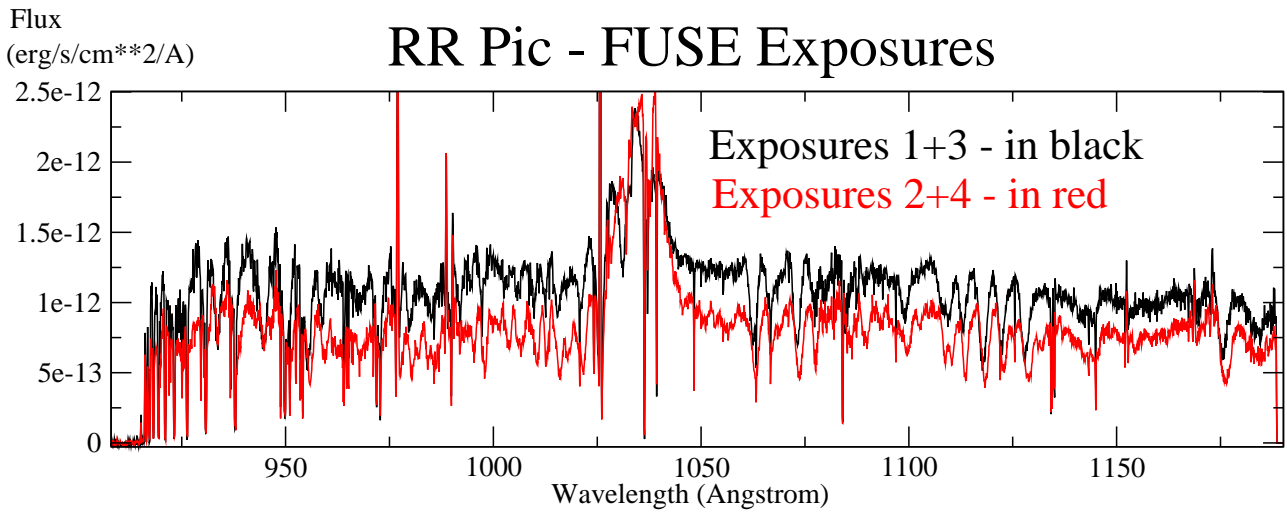} 
\caption{
The FUSE Spectrum of RR Pic is made of 4 FUSE exposures.
Exposures 1 and 3 are almost identical to each other, they  
have been co-added and the resulting spectrum is shown in black. 
Exposures 2 and 4 both show the same significant decrease in
flux and they too are almost identical to each other.
Exposures 2 and 4 have been co-added and are shown in red. 
} 
\end{figure}

\begin{figure}
\vspace{-5.cm} 
\plotone{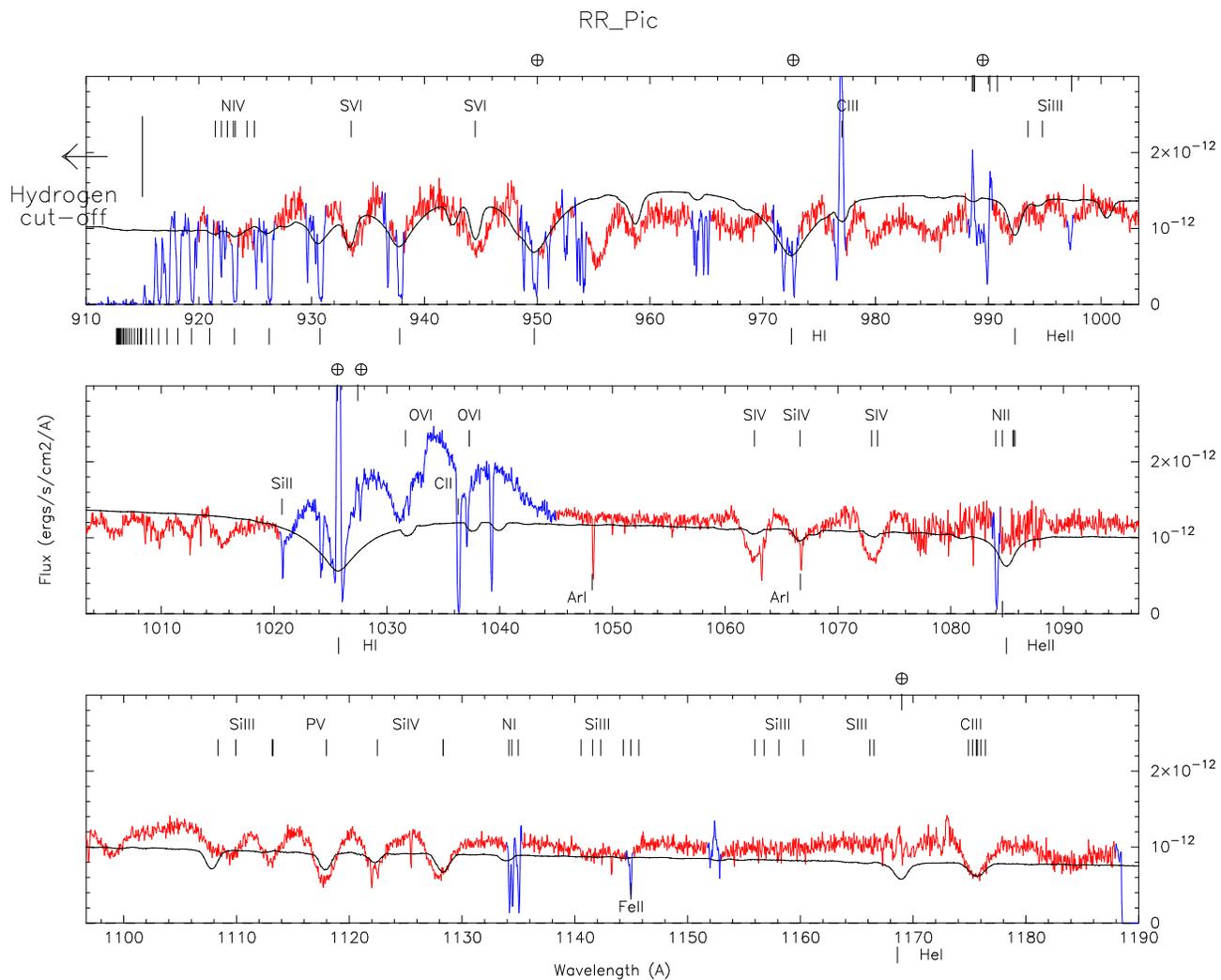} 
\vspace{1.cm} 
\caption{
The FUSE spectrum of RR Pic (in red) is modeled with a WD photosphere
(solid black line). 
The WD model has a temperature of 80,000K, with $\log(g)=7.0$, corresponding to 
a radius of 0.033$R_{\odot}$, giving a distance of 518pc. The high temperature
is needed to fit the shortest wavelength region of the spectrum, and the large
radius is derived from scaling the theoretical spectrum to the known distance. 
The known ISM absorption lines, the oxygen doublet
emission region, geocoronal emission lines as well as other
artifacts have been omitted from the spectrum and are marked
in blue.  
} 
\end{figure}

\begin{figure}
\vspace{-5.cm} 
\plotone{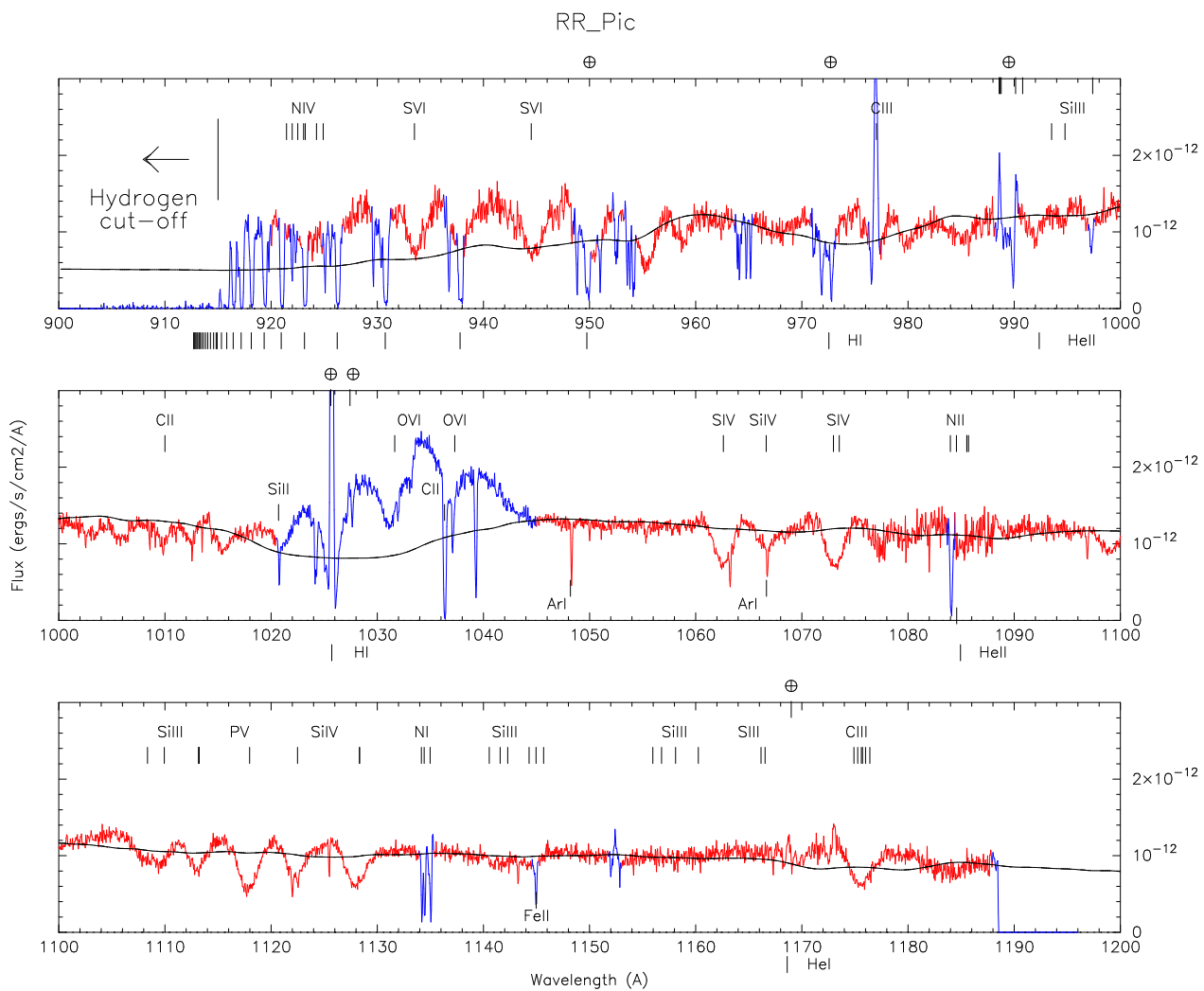} 
\caption{
Best Fitting model disk (in black) to the FUSE spectrum of RR Pic
(in red).
The  accretion disk model has an inclination of i = 60 degrees, 
a WD mass $M_{\rm wd} = 1.03 M_{\odot}$ 
and a high accretion rate of $10^{-8} M_{\odot}$/yr, giving a distance of 506pc, 
a result which is with the range of RR Pic's parallax distance.
The regions affected by the ISM and broad emission lines have been masked
before the fitting and are marked in blue. Note that the absorption lines of
the source are not masked as they were initially tentatively modeled 
with a hot WD model.   
} 
\end{figure}

\begin{figure}
\vspace{-5.cm} 
\plotone{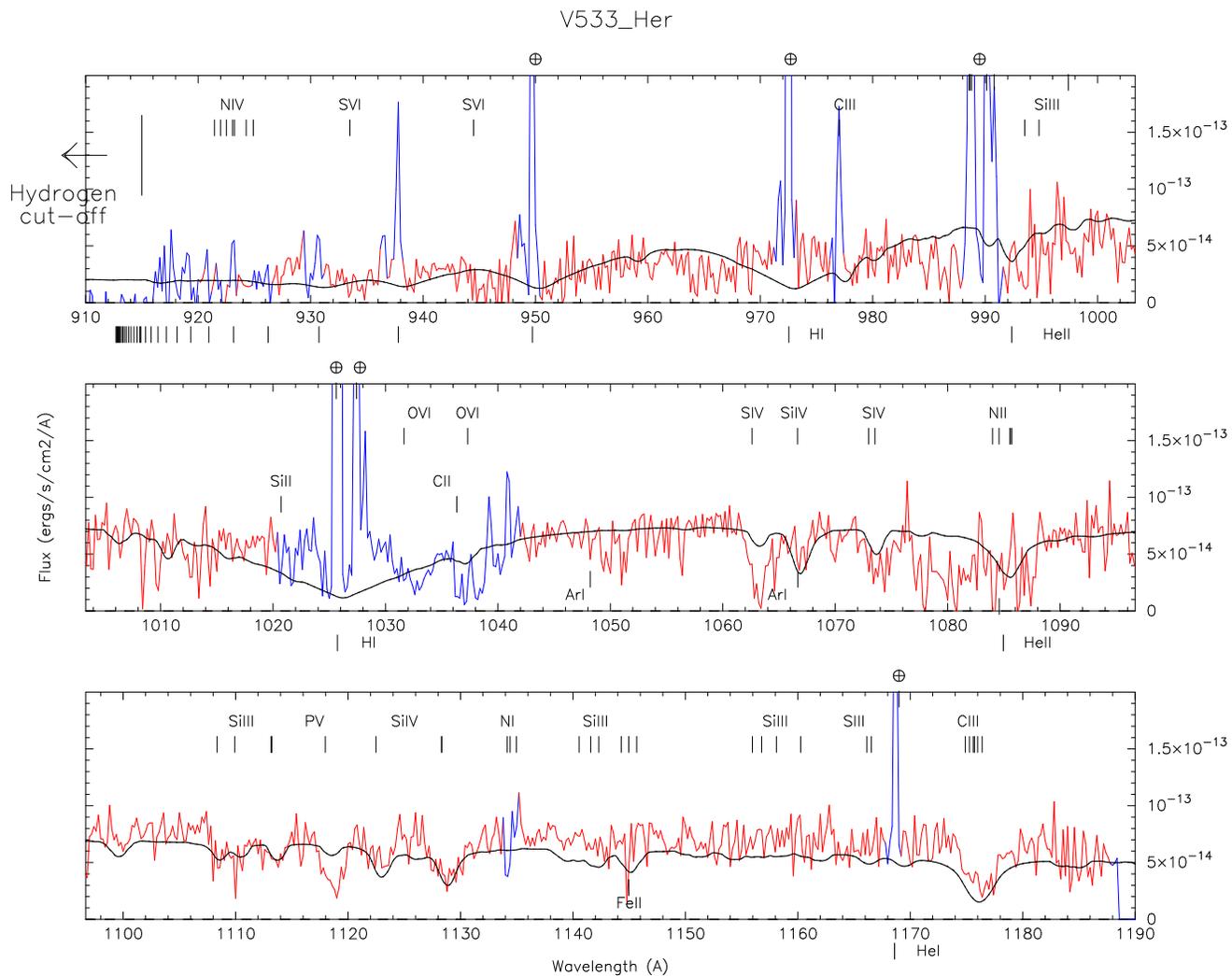} 
\vspace{1.cm} 
\caption{
Best Fitting WD solar abundance model to the FUSE spectrum of V533 Her. 
The WD temperature has been set to 40,000K, with $\log(g)=8.6$, 
corresponding to a $1 M_{\odot}$ WD with a radius of 6,000km,
and giving a distance of only     316pc, or about 1/2 to 1/4 the 
accepted distance.  
} 
\end{figure}

\begin{figure}
\vspace{-5.cm} 
\plotone{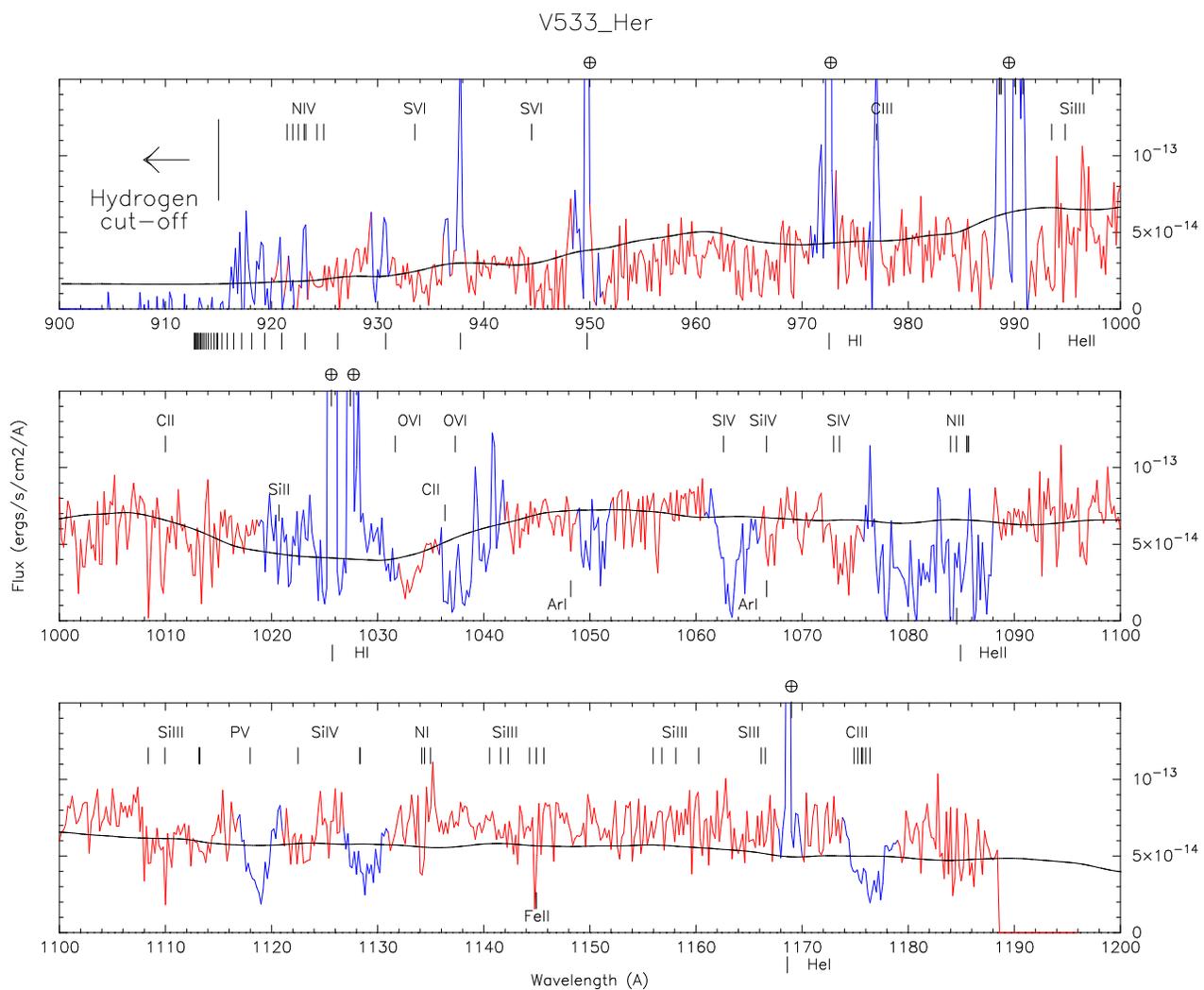} 
\caption{
Best Fitting model accretion disk to the FUSE spectrum of V533 Her 
for a $1.0 M_{\odot}$ WD with an accretion rate of $10^{-9}M_{\odot}$/yr and 
i = 60 degrees. 
This model provides a good continuum fit and a reasonable 
distance of 832 pc, but does not produce any absorption lines
because of the large Keplerian velocity broadening effect at
an angle of 60 degrees.  
} 
\end{figure}

\begin{figure}
\vspace{-5.cm} 
\plotone{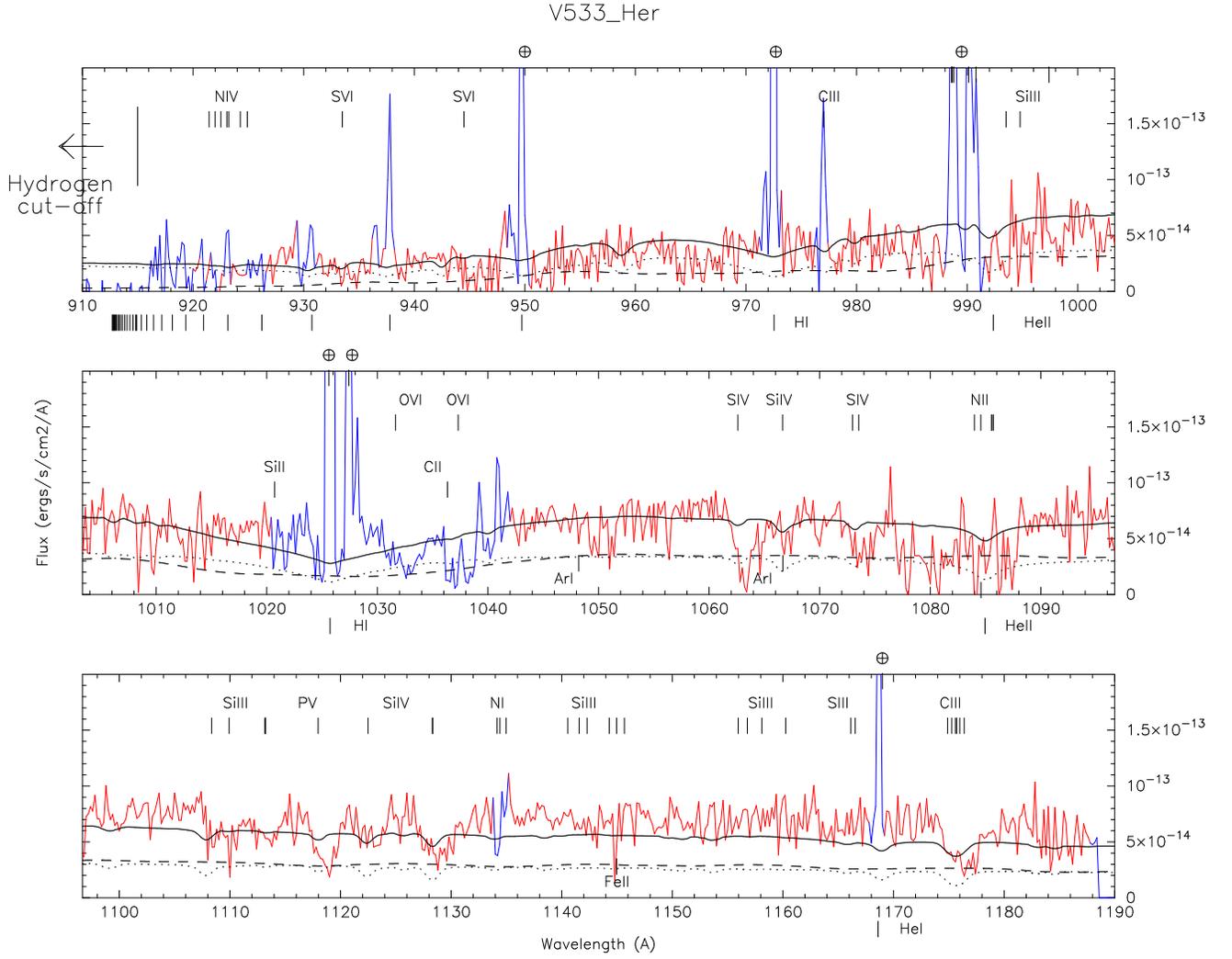} 
\vspace{1.cm} 
\caption{
Best Fitting accretion disk + WD model (solid black line) 
to the FUSE spectrum of V533 Her 
for a $1.0 M_{\odot}$ WD. The disk contributing 49\% of the FUV flux
(dashed black line) 
has an accretion rate of $10^{-9.5}M_{\odot}$/yr,  
and the WD contributing the remaining 51\% (dotted black line) has a 
temperature of 60,000K. The distance obtained from the fit is 640pc.  
} 
\end{figure}

\begin{figure}
\vspace{-5.cm} 
\plotone{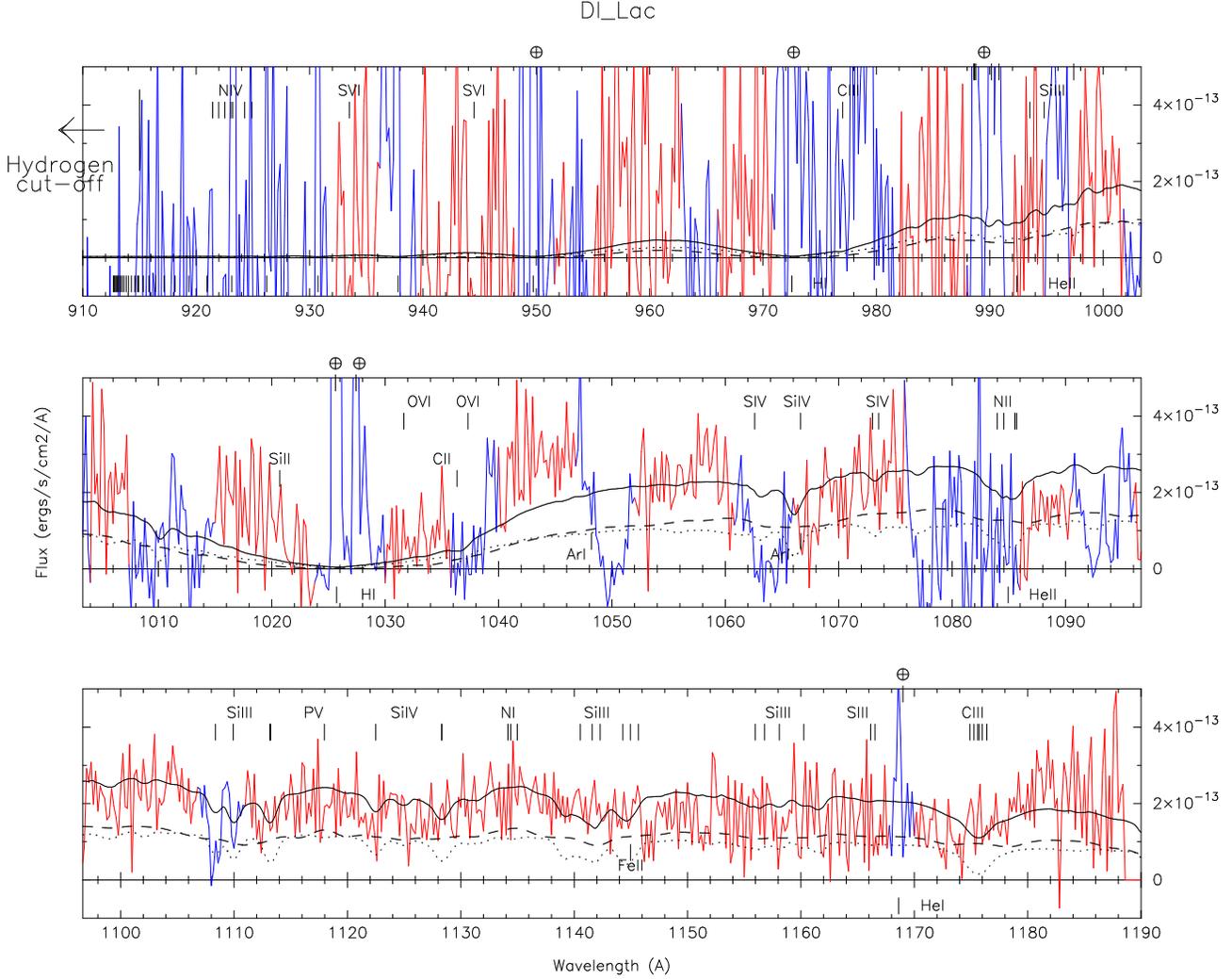} 
\caption{ 
Best Fitting accretion disk plus white dwarf synthetic spectrum (in solid
black)  to the combined FUSE + IUE spectrum of DI Lac (in red). 
The FUSE region is shown in details, the IUE region is shown in the
next figure. The regions of negative flux is also displayed to
show the very low quality of the data in the shorter wavelengths
(upper panel). 
The WD model (black dotted line) has a mass of $0.8M_{\odot}$ and 
a temperature of 30,000 K. The disk model (black dashed line) has 
a mass accretion rate of $\dot{M}=10^{-10}M_{\odot}$/yr and an inclination
of $i=18^{\circ}$. The distance obtained from the fit is 175 pc. 
Airglow emission lines and ISM absorption features have been masked (in blue) 
and are not modeled, only the red portion of the spectrum is modelled.  
}
\end{figure} 

\begin{figure}
\vspace{-25.cm} 
\plotone{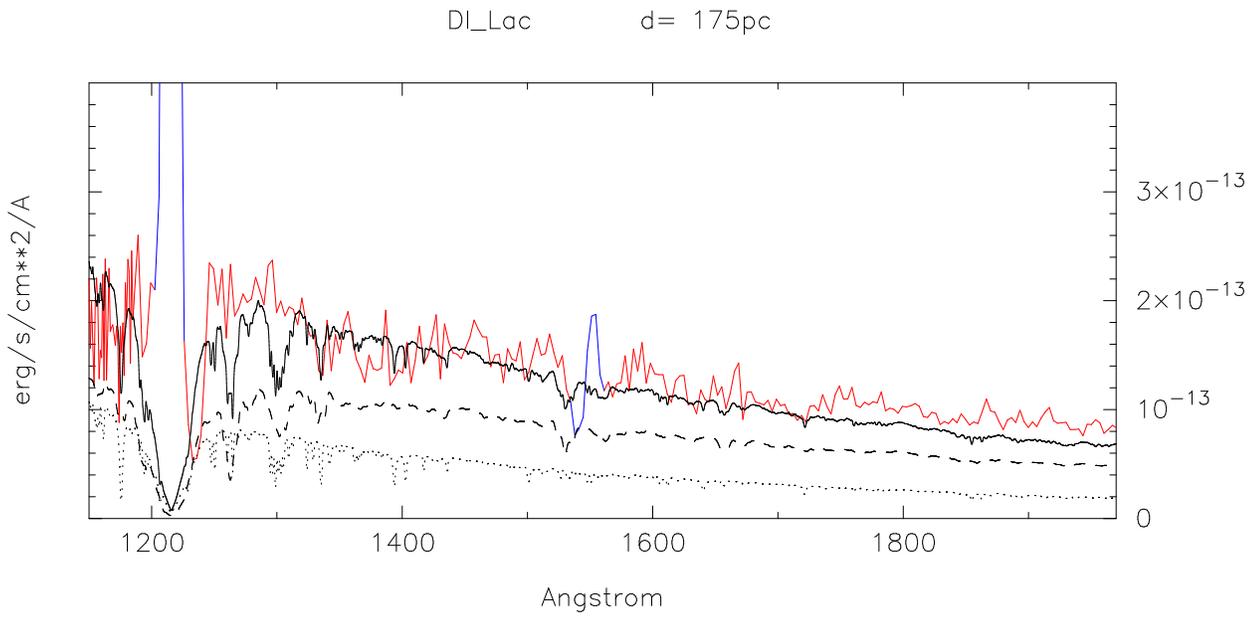} 
\caption{ 
Same as in Fig.11, but here the IUE region is shown. 
}
\end{figure} 

\end{document}